\documentclass[12pt]{article}

\usepackage{layout}
\usepackage{amsmath}
\usepackage{textcomp}
\usepackage{hyperref}

\newcommand{\clM}{\mathcal{ M}}

\newcommand{\clL}{{\cal L}}

\newcommand{\clE}{\mathcal{E}}

\newcommand{\rgl}{\rangle}
\newcommand{\lgl}{\langle}
\newcommand{\ep}{\epsilon}
\newcommand{\vep}{\varepsilon}
\newcommand{\tlG}{\tilde{G}}
\newcommand{\hclL}{\hat{\cal L}}
\newcommand{\clC}{{\cal C}}
\newcommand{\be}{\begin{equation}}
\newcommand{\ee}{\end{equation}}
\newcommand{\bea}{\begin{eqnarray}}
\newcommand{\eea}{\end{eqnarray}}

\begin{document}

\title{Electro-chemical manifestation of nanoplasmonics in fractal
media}


\author{Emmanuel~Baskin, Alexander~Iomin \\
\\
Department of Physics and Solid State Institute,
Technion, 32000, Haifa, Israel \\
\\ 
Cent. Eur. J. Phys.  \textbf{11}(6)  (2013)  676-684 }

\maketitle
\begin{abstract}
Electrodynamics of composite materials with fractal
geometry is studied in the framework of fractional calculus. This
consideration establishes a link between fractal geometry of the
media and fractional integro-differentiation. The photoconductivity in
the vicinity
of the electrode-electrolyte fractal interface is studied. The
methods of fractional calculus are employed to obtain an
analytical expression for the giant local enhancement of the
optical electric field inside the fractal composite structure at
the condition of the surface plasmon excitation. This approach
makes it possible to explain experimental data on
photoconductivity in the nano-electrochemistry.



\end{abstract}



\section{Introduction}

It is well known that small-particle composites, like
nano-structured noble-metal-dielectric composites, establishes a
striking response to the electromagnetic irradiation
\cite{Shalaev,Stockman,Stock1}. This effect differs radically from
one taking place in both ordinary bulk materials or individual
nano-particles. We  are speaking about so-called hot spots, which
result from strong localization of optical energy on the nanometer
scale (less than the optical wavelength). This effect is inherent
in disordered clusters of nano-particles at the condition of the
surface plasmon excitation.

This effect has a more pronounced manifestation on fractal
clusters. Fractals are not translationally invariant systems (as
any disordered systems), and, therefore, can not transmit running
waves \cite{Bunde}. As a result, dynamical excitations, like
surface plasmons, tend to be localized in fractals. Note, that
a superposition of electric fields induced by local surface plasmons
results in focusing and local giant enhancement of the electric field.
This giant enhancement has a geometrical nature of the
scale invariance of fractals, which, at some extend, has the the same nature
as X-ray diffractive maxima are the geometrical effect of a
spatial periodicity of crystals.

In this paper we employ  a concept of fractals and fractional
calculus for application to the electrolysis enhancement in
nano-electrochemistry. An enhancement of the electrolysis current
due to a rough metal-electrode interface was first observed in
1926 \cite{kaplan1}. The further enhancement can be also observed
when the electrolysis set up is extend to the light
electromagnetic field \cite{baskin,Cronin,Mukherjee}. This optical
induced current is due interplay between the light and fractional
nanostructure which is the double electric layer on the fractal
electrode interface. This leads to a local giant enhancement of
electric field. Technically, these ``hot spots'' lead to the
breakdown of the double layer capacitor that, eventually, changes
the electrode potential (potential of the Stern layer). The hot
spot phenomenon is well known in optics, and relates to optical
properties of metal nanoparticles, and is largely due to their
ability to produce giant and highly localized electromagnetic
fields. Two phenomena, closely related to each other, are
responsible for giant enhancement. The first one is localized
surface plasmon resonance (see, for examples the reviews
\cite{Stock1,Stockman}, or monograph \cite{Shalaev}), which is
charge density oscillations confined to conducting (metallic)
nanostructures. Strong enhancement of the electromagnetic field is
possible if its frequency is near the frequency of these
oscillations \cite{Shalaev,Stockman,st1,st2,st3}. Another giant
enhancement due to fractal geometry is a so-called geometrical
enhancement \cite{bi2011b}, obtained analytically in the framework
of fractional calculus, where a question of its experimental
manifestation is still open \cite{zbi2013} .

Application of fractional integro-differentiation and its link to
fractals and fractal geometry may shed a new light on studies of
properties of composite materials with underlying fractal
structures \cite{mandelbrot,gouyet,west90,benavraam}. The fractal
concept makes it possible to involve the application of the
powerful methods of fractional integro-differentiation, see
\textit{e.g.}, \cite{podlubny,oldham}. Extended reviews on
fractional calculus can be found in \cite{podlubny,oldham,SKM},
and a brief survey is presented in Appendix A.

In this paper, we suggest an analytical consideration, based on
fractional calculus, for the purpose of obtaining an analytical
expression for the electric field in fractal composite media,
namely in fractal electrode-electrolyte interface. This approach
establishes a relation between the fractal geometry of the medium
and fractional integro-differentiation, and it is also based on
our previous studies \cite{bi2011b,bi2011a,bi2012}. We suggest a
coarse graining procedure for the electric field in the Maxwell
equation to treat the electric induction term, which is a
discontinuous function in the fractional composite media. This
smoothing procedure makes it possible to involve the fractional
calculus in (quasi-)continuous media \cite{RenNig} and, in the
present particular case, to obtain an equation for the electric
field in a closed form including the fractal polarization charge
distribution. The main idea is due a seminal result \cite{NIG92},
where a link between fractal geometry and fractional
integro-differentiation is constituted in the procedure of
averaging an extensive physical value. It should be admitted that
this description of properties of fractal media in the framework
of the fractional calculus is possible only in some approximation.
This procedure of filtering, expressed by means of a smooth
function over a Cantor set, leads to fractional integration. In
its eventual form, it has been presented in Ref. \cite{NIG98}.
This topics was extensively studied \cite{RenNig,NIG92} in
application for signal processing and dielectric relaxation in
complex media \cite{Osokin1,Osokin2,Nigmatulin}.

The main idea of filtering, or embedding a matter inside a fractal
is the construction of a convolution integral according
integration over the fractal volume $\mu(x)\sim x^{\alpha}$
\be\label{fpe3} %
\lgl{F(x)}\rgl=\int F(x)d\mu(x)\Rightarrow {}_0I_x^{\alpha}F(x) \,
, \ee %
where ${}_0I_x^{\alpha}$ designates the fractional integral of the
order of $\alpha$ in the range from $0$ to $x$:
\be\label{fpe4}   %
{}_0I_x^{\alpha}F(x)=\frac{1}{\Gamma(\alpha)}
\int_0^x(x-x')^{\alpha-1}F(x')dx'\,
.  \ee   %
Here $\Gamma(\alpha)$ is a gamma function and $\alpha$ is the fractal dimension.
For example, in Ref.
\cite{NIG98} this kind of integration has been obtained by means
of averaging over one of a parameter, which describes a fractal
structure. Considering a convolution integral $ F(x)=W(x)\star
f(x)=\int_0^xW(x-y)f(y)dy$ with the function $W(x)$, which obeys
the scaling relation $W(x)=\frac{1}{a}W(bx)$ and describes a
Cantor set, one performs averaging over the parameter $b$ (namely
$\ln b$ \cite{NIG98}) and obtains Eq. (\ref{fpe3}), where the
averaging reads $ \lgl{F(x)}\rgl_{\ln b}=A(\alpha,b)
{}_0I_x^{\alpha}f(x)$, where $A(\alpha,b)$ is a constant factor
and $\alpha=\frac{\ln a}{\ln b}$. Instead parameters $a$ and $b$
one introduces the fractal dimension $\alpha$ and the log-periodicity
parameter $\ln b$. Therefore, averaging over the log-periodicity means
averaging over the all possible realizations of the fractal.

It should be stressed that this idea was firstly expressed by
Kolmogorov in \cite{kolmogorov} for fractional Brownian motion,
and later rediscovered by Mandelbrot and Van Ness, who explored
this phenomenon in greater detail  in \cite{mandelvanness}.

%

As shown \cite{bi2011b,bi2012,bi2011a}, these mathematical
constructions are also relevant to the study of electrostatics of
real composite structures. The main objective of the present
research is to solve a standard electrostatic problem, namely, the
derivation of the electric field in a fractal metal-dielectric
composite on the electrode-electrolyte interface in the presence
of the double layer capacitor. This leads to the exploration of
the fractional Maxwell equations. This issue attracts much
attention and is also extensively studied to explore the role of
fractional calculus in electrodynamics
\cite{engheta1997,Naqvi,naber,tar2005,tar20052008} (see review
\cite{tarasov2008}).

This formulation can be also employed for the
fractal geometry, in particular, for the exploration of the
electric field in fractal dielectric composites
\cite{epje2012,bi2012}, or metal--dielectric composites
\cite{bi2011b,bi2012}. This problem is important for
nanoplasmonics, where the interplay between the light and
fractional nanostructures leads to a local giant enhancement of
the electric field \cite{Shalaev,Stockman}.

\section{Electro-mechanical consideration of electron motion in the double
layer}

The double layer plays a central role for stability in
electrolysis current phenomenon near the electrodes, see, for
example, \cite{Butt}. An electron motion inside the double layer
in the presence of the external high-frequency electric field is
described by equation
\begin{equation}\label{LL1}
m\ddot{x}=-\frac{dU}{dx}+f(\omega t)\, .
\end{equation}
Here $U(x)$ is the double layer constant field potential, which is
considered as a capacitor potential, while
$f(\omega t)=f_0\cos\omega t=eE_0(t)$
is a fast-oscillating  perturbation, such that
$\omega\mathcal{T}\gg 1$, where $\mathcal{T}$ is an averaged time
propagation through the double layer. Here $E_0(t)$ is the
electric component of the external light and $e$ the electron
charge. It is well known that the average procedure over the fast
oscillations leads to the additional term to the potential due to
the kinetic term \cite{LLI}
\begin{equation}\label{LL2}
U_{\rm eff}=U+\frac{f_0^2}{4m\omega^2}=U+U_f\, .
\end{equation}

This additional term leads, eventually, to effective decrease of
the double layer capacitor potential $W\rightarrow
We^{-\frac{U_f}{kT}}$, where $k_BT$ is the Boltzman temperature.
To show this let us consider the Poisson-Boltzmann theory of the
diffusive double layer \cite{Butt}.

\subsection{The Poisson-Boltzmann equation}

Now we follow the Poisson-Boltzmann theory near the double layer
\cite{Butt}. For the considered planar surface, the $x$ coordinate
determines the direction of the potential change in the double
layer. Therefore, the electric potential $\psi(x)$ near a charged
planar electrode interface with the surface potential
$\psi(x=0)=\psi_0$, taken as the boundary condition for the double
layer, is determined by the one-dimensional Poisson equation
\begin{equation}\label{Pois1}
\frac{d^2\psi(x)}{d\,x^2}=-\frac{\rho(x)}{\vep_d}\, ,
\end{equation}
where $\rho(x)$ is the local electric charge density and $\vep_d$
is the permittivity of dielectric solution (at this point it does
not carry any important information). The charge density is
determined by the Boltzmann statistics that reads for the local
ion density $n_i=n_0e^{-A_i/k_{B}T}$, where $A_i$ is the performed
work to bring an ion from infinity to the point $x$ inside the
double layer. In the presence of the additional potential term
$U_f$ in (\ref{LL2}) due to breakdown of the double layer
capacitor, the electric work $A_i$ depends on the ion charge.
Namely, to bring cation the required work is
$A_{-}-U_f=-e\psi(x)$, while for anion it is $A_{+}+U_f=e\psi$.
Therefore, the effective potential decreases, and since $U_f$ is
independent of the coordinates $x$, this leads to only to
re-scaling of the boundary condition
$psi_0\rightarrow\psi_0-U_f/e$. This leads to change of the
constant value of the solution of the Poisson-Boltzman equation
(\ref{Pois1}) $\psi(x)=(\psi_0-U_f/e)\Psi(x)$, where $\Psi(x=0)=1$.

To see this effect of the external light influence on the
electrolysis, one needs $U_f\sim k_{B}T$. If the electric
component of the external light, with $\omega\sim 10^{15}$, is
$E_0\sim 1\div 10\, V/cm$, the additional potential energy is
$U_f\sim (10^{-9}\div 10^{-7})k_{B}T$, which is vanishingly
smaller than temperature fluctuations. Therefore, a condition of
observation this effect of the influence of the external light on
the electrode potential is an essential
enhancement of the electric field of the order of $10^5\div 10^6$
times due to the surface plasmon resonance that yields the
respond/enhanced electric field of the order of $10^6\div
10^7$(V/cm). Effect of this giant enhancement of the external
low-intensity light is considered in the quasi-static
approximation.

\section{Quasi-static Electromagnetic Field}

The fractal nanostructure placed on the electrode surface is
embedded in the $3D$ space, therefore, we consider a sample of a
size $R$ which consists of a fractal metal nanosystem embedded in
a dielectric (electrolyte) host medium. The system is subject to
an external electromagnetic field $\mathbf{E}_0(t)$ at an optical
frequency $\omega$, and in the experimental setup, the double
layer fractal-host inhomogeneities (of the order of $R$) are much
smaller than the light wavelength $\lambda$. In this case, a
quasi-static approximation is valid \cite{LLVIII}. Therefore, the
nanosystem contains a metallic fractal with a fractal volume
$V_D\sim R^D$ and the permittivity $\vep_m(\omega)$, which depends
on optical frequency $\omega$, while the dielectric-electrolyte
host has the volume $V_h$ and the permittivity $\vep_d$. At the
considered scale, the magnetic component is not important and
cannot be considered. The electric component satisfies the Maxwell
equation in the Fourier frequency domain
\be\label{qsef1}   %
\nabla\cdot\Big[\vep(\mathbf{r},\omega)\mathbf{E}(\mathbf{r},\omega)\Big]=0\, . \ee   %
The permittivity can be expressed by means of a characteristic
function $\chi(\mathbf{r})$ \cite{st3,st4,bi2011b} in the form
$\vep(\mathbf{r},\omega)=\vep_m(\omega)\chi(\mathbf{r})+\vep_d[1-\chi(\mathbf{r})]$,
where the characteristic function inside the fractal is
$\chi(\mathbf{r})=1$, $\mathbf{r}(x,y,z)\in V_D$, while inside the
dielectric host it reads $\chi(\mathbf{r})=0$,
$\mathbf{r}(x,y,z)\in V_h$. Following Ref.~\cite{bi2011b}, one
splits the electric field into the two components $
\mathbf{E}(\mathbf{r})=\tilde{\mathbf{E}}(\mathbf{r})+\mathbf{E}_1(\mathbf{r})$
and looks for the reaction of the nanosystem on the external
field, where $\tilde{\mathbf{E}}(\mathbf{r})$ is the electric
field induced by $\mathbf{E}_0$ in the homogeneous nanosystem,
when $\chi(\mathbf{r})=0$ for $\forall ~ \mathbf{r}$, or
$\chi(\mathbf{r})=1$ for $\forall ~ \mathbf{r}$ and
$\nabla\tilde{\mathbf{E}}(\mathbf{r})=0$. Here
$\mathbf{E}_1(\mathbf{r})$ results from inhomogeneity of the
nanostructute. Thus Eq. (\ref{qsef1}) can be rewritten in the form
\be\label{qsef1_a}  %
\chi(\mathbf{r})\nabla\cdot\mathbf{E}_1(\mathbf{r})-q(\omega)
\nabla\cdot\mathbf{E}_1(\mathbf{r})+
\mathbf{E}_1(\mathbf{r})\cdot\nabla\chi(\mathbf{r}) =
-\tilde{\mathbf{E}}(\mathbf{r})\cdot\nabla\chi(\mathbf{r})\, , \ee   %
where $q(\omega)=\frac{\vep_d}{\vep_d-\vep_m(\omega)}$. In general
case, mixed (Dirichlet, or Neumann) boundary conditions are
imposed \cite{st3,st4}. It follows from the boundary conditions
that $\nabla\cdot \mathbf{E}(\mathbf{r},\omega)=0$ with
$\mathbf{E}(\mathbf{r}\in S,\omega)=\mathbf{E}_0(\omega)$, where
$S$ means boundaries. This also supposes $\chi(\mathbf{r}\in
S)=0$.

\subsection{Filtering inside a random fractal}

Since the characteristic function $\chi(\mathbf{r})$ is
discontinuous, the next step of our consideration is
coarse-graining the electric field, which is averaging Maxwell's
Eq. (\ref{qsef1_a}). This procedure relates to integration with
the characteristic function
$\int\chi(\mathbf{r})\nabla\mathbf{E}_1(\mathbf{r})dV$ and
evaluation of the integral of the polarization charge density term
$\int\nabla\chi(\mathbf{r})\cdot\mathbf{E}(\mathbf{r})dV$. To that
end, let us consider a spherical volume of the radius $r$,  such
that $l_0\ll r< R$, where $l_0$ is a minimal size of the
self-similarity of the fractal volume on the electrode
corrugated/fractal surface. In the sequel we will work with
dimensionless variable $r/l_0\rightarrow r$. The electric field
does not change at this scaling. A fractal mass inside the volume
is $\clM(r)\sim r^D$, where $0<D<3$. Therefore, an average density
of the metallic phase is of the order of $r^{D-3}$. For the
filtering mass inside the fractal, we consider convolution in Eqs.
(\ref{fpe3}) and (\ref{fpe4}). We also, reasonably, suppose that
the random fractal composite is isotropic, \textit{i.e.}, the
(averaged) similarity exponents coincides along all directions.
This yields that one takes into account only radius $r$ and any
changes in the inclination and azimuth angle directions can be
neglected (this means that the averaged fractal aggregate is
considered as a sphere with a radially dependent filling factor
$r^D$). Then, we have for the divergence
\be\label{qfsef2}  %
\nabla\cdot\mathbf{E}_1(\mathbf{r})=
\nabla_rE_{r,1}(r)\equiv\nabla_rE_1(r) \, . \ee   %
Using property of the characteristic function, we obtain that the
measure of the fractal volume $V_D$ is
$\mu(V_D)=\int\chi(V_D)dV=\int\rho dV_D$, where
$dV_D=\frac{2^{3-D}\Gamma(3/2)}{\Gamma(D/2)}|\mathbf{r}|^{D-3}r^2dr\sin\theta
d\theta d\phi $ is the elementary fractional volume in the
spherical coordinates and $\rho$ is a fractal density factor. In
what follows, we will use $\mu(r)$ as the fractal volume by means
of the fractal density $\sum_{r_j\in V_D} \delta(r'-r_j)$ and for
simplicity, violate all constants. Therefore, the fractal volume
reads
\be\label{qfse3a}  %
\mu(r)=r^D=\int_0^r\sum_{r_j\in V_D} \delta(r'-r_j){r'}^2dr'\,
. \ee  %
Now, filtering inside the fractal due to characteristic function,
which depends only on the radius $r$, yields the integrations
\be\label{qsef3}   %
\frac{1}{4\pi}\int\chi(\mathbf{r})\nabla_rE_1(r)dV =
\int_0^r\chi(r')[\nabla_rE_1(r')]{r'}^2dr'
=\int_0^r\sum_{r_j\in V_D} \delta(r'-r_j)G'(r')dr' \ee %
Here we define $G(r)=r^2E_1(r)$ and $G'(r)\equiv \frac{d}{d\,r}G$.

Following Ref. \cite{Ren} (Theorem $3.1$), we obtain
\be\label{theorRen}  %
\int_0^rG'(r')d\mu(r')\sim\frac{1}{\Gamma(D-2)}
\int_0^r(r-r')^{D-3}G'(r')dr'
\equiv {}_0I_r^{D-2}G'(r)\, . \ee %
Therefore, we consider the integration in Eq. (\ref{qsef3}) as the
convolution integral with the averaged fractal density
$(r-r')^{D-3}$.

\paragraph{Polarization term}

Now we estimate the integral
$\int_0^rE(r')\nabla_{r'}\chi(r'){r'}^2dr'$. The fractal dust
$V_{D}$ at the $N$th step of the construction consists of balls
$B_N$ with  the radius $\Delta_N$. For example, $\Delta_N\sim
l_0$. In the limiting case one obtains
$V_{D}=\lim_{N\to\infty}\bigcup B_N$ \cite{falconer}. The
characteristic function for every ball is $\chi(\Delta_N)=
\Theta(r-r_j)-\Theta(r-r_j-\Delta_N)$. Differentiation of the
characteristic function on the intervals $[r_j,r_j+\Delta_N]$
yields $ \nabla_r\chi(\Delta_N)
=\delta(r-r_j)-\delta(r-r_j-\Delta_N)$. Therefore, for any
interval $\Delta_N$ and at $r=r_j$, integration with the electric
field yields
$${}_rI_{r+\Delta_N}^1E(r)r^2\nabla_r\chi(r)
=E(r)r^2-E(r+\Delta_N)(r+\Delta_N)^2\, . $$   %
This expression is not zero in the limit $\Delta_N\rightarrow 0$.
Let $E(r_j)$ is the electric field outside the ball $B_N$ and
$E(r_j+\Delta_N)$ denotes the internal electric field. The
relation between them, due to Eq. (8.2) in Ref.~\cite{LLVIII} for
polarization of a dielectric ball, is
$$E(r_j+\Delta_N)=\frac{3\vep_d}{\vep_m(\omega)+2\vep_d}E(r_j)\, .
$$  
Therefore, the shift for the electric field in the limit
$\Delta_N\rightarrow 0$ is
\be\label{qsef4}  %
E(r_j)-E(r_j+0)=E(r_j)\frac{\vep_m(\omega)-
\vep_d}{\vep_m(\omega)+2\vep_d}\, . \ee    %
Finally, integration of the polarization charge term yields
\be\label{qsef5} %
\int_0^rE(r')\nabla_{r'}\chi(r')r'^2dr'\equiv
{}_0I_{r}^1E(r)r^2\nabla_r\chi(r) =\frac{\vep_m(\omega)-
\vep_d}{\vep_m(\omega)+2\vep_d}\sum_{r_j\in
V_D}\int_0^rE(r'){r'}^2\delta(r'-r_j)dr'\, . \ee %
Again, one obtains the integration of the electric field with the
fractal density $\mu(r')=\sum_{r_j\in V_D} \delta(r'-r_j)$. This
corresponds to the fractal volume (\ref{qfse3a}), and, hence, we
consider the integration in Eq. (\ref{qsef5}) as the convolution
integral of Eq. (\ref{theorRen})  with the averaged fractal
density $(r-r')^{D-3}$. Note that this procedure is exactly
corresponds to Nigmatulin's arguments on filtering according Eq.
(\ref{fpe3}), where the convolution integral exists from the
beginning, while in our case, the convolution integral appears
owing to theorem of Ref. \cite{Ren} for integration with the
fractal volume/measure $\mu(r)$. Finally, commenting this
procedure of filtering inside the fractal, it should be stressed
that integration with the fractal characteristic function in Eq.
(\ref{qsef3}) leads to the integration with the fractal measure
$\int\chi(r)G'(r)dr=\int G'(r)d\mu(r)$ that results from the
physical properties of $\chi(r)$. On the contrary, integration
with the differential $d\chi(r)$ in Eq. (\ref{qsef5}) leads again
to the non-zero integration with the fractal measure
$\int\chi'(r)G(r)dr=\int G(r)d\mu(r)$ due to the properties of the
electric field, namely boundary properties of the electric field
on the rough metal-electrode-electrolyte interface in the double
layer capacitor.

Eventually, we obtain for the polarization charge term in Eq.
(\ref{qsef1_a})
\be\label{qsef6}   %
\int_0^rE(r')\nabla_{r'}\chi(r')r'^2dr'\sim
-\frac{p(\omega)}{\Gamma(D-2)}\int_0^r(r-r')^{D-3}E(r'){r'}^2dr'
\equiv {}_0I_r^{D-2}[G(r)+\tilde{E}_{0}r^2]\,
. \ee   %
Here $\tilde{E}_0=\tilde{\mathbf{E}}\cdot\hat{\mathbf{r}}$ is a
projection of the external electric field on the radial direction
inside the chosen spherical volume of the radius $r$ and
\be\label{p_omega}  %
p(\omega)=\frac{\vep_d-\vep_m(\omega)}{\vep_m(\omega)+2\vep_d}
\, .  \ee  %

Taking all these arguments and results in Eqs. (\ref{theorRen})
and (\ref{qsef6}), one presents Eq. (\ref{qsef1_a}) in the
coarse-graining form
\be\label{qsef7}   %
{}_0I_r^{D-2}G'(r)-q(\omega){}_0I_r^1G'(r)-
p(\omega){}_0I_r^{D-2}G(r)
= \frac{2p(\omega)\tilde{E}_0}{\Gamma(D+1)}r^D\, .\ee %
The Laplace transform can be applied to Eq. (\ref{qsef7}). Since
$G(r=0)=G'(r=0)=0$ (the electric field of fractal charge density
diverges slowly than $\frac{1}{r^2}$
\cite{bi2011b,bi2011a,tar2005}), one obtains for
$\tlG(s)=\hclL[G(r)]$ due to Eq. (\ref{mt3}) (see Appendix):
\be\label{qsef8}   %
\tlG(s)=\frac{2p(\omega)\tilde{E}_0}{\vep_ds^{3}}
\cdot\frac{1}{s-q(\omega)s^{D-2}-p(\omega)}\, .  \ee  %
Note that the second term in Eq. (\ref{qsef7}) is $q(\omega)G(r)$.

Before arriving at the main result, let us consider the limiting
cases. For $\vep_m(\omega)=\vep_d$ one obtains $q(\omega)=\infty$
and $p(\omega)=0$. This yields $E_1(r)=0$, and the solution for
the electric field is $E(r)=\tilde{E}_0\equiv\tilde{E}_r$. Another
limiting case is $|\vep_m(\omega)|\rightarrow\infty$. In this case
$q(\omega)=0$ and $p(\omega)=-1$, thus $E_1\sim \tilde{E}_0$.
Important result here is that permittivity of the mixture is
approximately $\vep_d$ that corresponds to the well known result
in Ref.~\cite{LLVIII} [see Eq. (9.7)].

\section{Surface plasmon resonance}

Now we consider a condition of the strong enhancement of the
electric field at the surface plasmon resonance,  when
$Re[\vep(\omega)]=-2\vep_d$. This also known as the Fr\"ohlich
resonance \cite{Frohlich,bohren}. The permittivity of the metallic
nanostructure inclusion at the resonance condition is a complex
value $\vep_m(\omega)=\vep_1+i\vep_2$, where $\vep_2/\vep_1\ll 1$
that is described by classical Drude formula (see \textit{e.g.}
\cite{LLVIII,bohren})
$\vep_m(\omega)=\ep_0-\frac{\omega_p^2}{\omega(\omega+i\gamma)}$,
where $\omega_p$ is a so-called plasma frequency, $\ep_0$ is a
high-frequency lattice dielectric constant, while the attenuation
coefficient $\gamma$ is small in comparison with the resonant
frequency. Therefore, we have for the metallic nanostructure
$\vep_m(\omega)=\vep_1+i\vep_2$, where $\vep_1={\rm
Re}[\vep_m(\omega)]=\ep_0-\omega_p^2/\omega^2$ and $\vep_2={\rm
Im}[\vep_m(\omega)]=\gamma\omega_p^2/\omega^3$.

At a small detuning from  the resonance, when
$Re[\vep(\omega)]=-2\vep_d+\vep_2$ that corresponds to the width
of the resonance in the frequency domain, $p(\omega)$ reaches the
maximal values, that yields
\be\label{spr1a}  %
p(\omega)=-1
+\frac{3\vep_d}{2\vep_2}(1-i)\approx\frac{3\vep_d}{2\vep_2}(1-i)\,
, \ee %
\be\label{spr1b}  %
q(\omega)=\frac{1}{3}-\frac{\vep_2}{9\vep_d}(1-i)
\approx \frac{1}{3}\, . \ee  %
These expressions are inserted in Eq. (\ref{qsef8}). Before
carrying out the inverse Laplace transform $\hclL^{-1}[\tlG(s)]$,
it is reasonable to simplify the second denominator. We recast Eq.
(\ref{qsef8}) in the form
\be\label{spr2}  %
\tlG(s)=2p(\omega)\tilde{E}_0\sum_{k=0}^{\infty}
\frac{[q(\omega)s^{D-2}+p(\omega)]^k}{s^{k+4}}\, .  \ee  %
We take into account that for the scale $r\gg 1$, the Laplace
parameter is small $s\ll 1$, and the binomial becomes
approximately a monomial. The Laplace inversion of Eq.
(\ref{spr2}) can be performed using an expression for the
Mittag-Leffler function \cite{BE}
\be\label{spr3}  %
\clE_{(\nu,\beta)}(zr^\nu)= \frac{r^{1-\beta}}{2\pi
i}\int_{\clC}\frac{s^{\nu-\beta}e^{sr}}{s^{\nu}-z}ds=
\frac{r^{1-\beta}}{2\pi}\int_{\clC}e^{sr}\sum_{k=0}^{\infty}
\frac{z^k}{s^{\nu k+\beta}}ds
=\sum_{k=0}^{\infty}\frac{[zr^{\nu}]^k}{\Gamma(\nu k+\beta)}\, , \ee %
where $\clC$ is a suitable contour of integration, starting and
finishing at $-\infty$ and $\nu,\beta>0$.  Comparing Eqs.
(\ref{spr2}) and (\ref{spr3}), $\beta=4$ and $\nu=1$, one obtains
for the electric field
\be\label{spr4}    %
E_1(r)=2p(\omega)\tilde{E}_0r\clE_{(1,4)}[p(\omega)r]\, . \ee  %
Since the argument of the Mittag-Leffler function is large, its
asymptotic behavior is \cite{BE}
\be\label{spr5}  %
E_1(r)\sim \frac{2\tilde{E}_0r}{p^2(\omega)}e^{p(\omega)r}\propto
\frac{\tilde{E}_0}{\vep_d}\exp\left[r\frac{3\vep_d}{2\vep_2}(1-i)\right]\,
.\ee   %
Eventually, we arrived at the exponential (geometrical)
enhancement and giant oscillations of the respond electric field
due to the fractal geometry of the metal-dielectric composite.

\subsection{Discussion on the geometrical enhancement of the
electric field}

As shown, the resulting (enhanced) electric field depends on the
parameter $p(\omega)$, which is the pre-factor and the argument of
the Mittag-Leffler function (when $r\sim 1$) in Eq. (\ref{spr4}),
and defined in Eq. (\ref{p_omega}). This parameter is absorbtion
efficiency in the electrostatic (quasi-static) approximation
\cite{bohren}. In our case, it describes polarization, and it is
obtained in Eq. (\ref{qsef5}) under evaluation of fractal boundary
conditions. One should recognize that the linear (classical
\cite{bohren}) enhancement of the electric field is always takes
place, due to Eq. (\ref{qsef5}) that corresponds to the
enhancement of the electric field by a small particle and that is
reflected by the pre-factor in Eq. (\ref{spr2}). For a fractal
small composite of many particles the situation differs
essentially. Here the geometrical enhancement of the electric
field is due to the focusing effect of fractal clustering
reflected by the Mittag-Leffler function and it depends on the
argument of the complex value of $p(\omega)$.

The exponential (geometric) enhancement takes place when
$|p(\omega)|\gg 1$ and ${\rm Re} p(\omega)>0$ as in Eq.
(\ref{spr4}). This condition is fulfilled for those frequencies
$\omega$ that are in the vicinity of the SPR: ${\rm
Re}p(\omega)=-2\vep_d+\Delta$, where $\Delta\sim\vep_2\ll \vep_d$.

We have to admit that the obtained expressions in the exponential
forms are the upper bound of the electric field enhancement. The
enhanced $E_1$ does not exceed of the order of $10^8$(V/cm),
otherwise the nonlinear effects become important that violates the
linear quasi-static consideration. This restriction yields
$r(\vep_d/2\vep_2)\leq 20$, which is a reasonable value for
experimental realizations. Therefore, we have the light wavelength
$\lambda\sim 10^{-4}$cm, fractal inhomogeneity size $l_0\sim
(10^{-6}$cm, and $\vep_d/\vep_2\sim\omega\tau\sim
5\div 15$, where $\omega$ is the optical frequency, while $\tau$
is the relaxation time. The latter value determines $l_0$, as
well, which was introduced above as a minimum self-similarity
size. Note that $\tau\leq \tau_s\equiv\frac{l_0}{v_F}$, where
$\tau_s$ is the surface relaxation time, while $v_F$ is the
velocity on the Fermi surface (for free electrons). Therefore,
from the condition $\omega\tau\gg 1$, one obtains $l_0>
\frac{v_f}{\omega}\sim 10^{-7}$cm.

\section{Conclusion}

It should be admitted that a traditional consideration of
photocatalysis is based on the energy loss due to plasmon's decay
that leads to generation of hot electrons and holes and suitable
increase of oxidation-reduction processes. This effect is usually
considered under investigation of macroscopic, electrochemical
manifestation of plasmon resonance \cite{Stock1,Louis,Knight}. The
main reason for this follows from the consideration of an
individual nanoparticle that yields the maximum enhancement of the
external field of the order of $10^2$. Therefore, there is no any
measurable chemical effects under such high frequency field
($\omega\sim 10^{15}sec^{-1}$) with the amplitude of the order of
$10^2$(V/cm), particularly under condition that this field exists
only in the vicinity of the nanoparticle surface (on the distance
of the order of the particle radius).

Following our consideration in Sec. 2, we were able to show that the
manifestation of breakdowns of Stern double layer-capacitor due to
the hot spots possible at the enhanced electric field of the order
of $10^6\div 10^7$(V/cm), and the observed effect is due to the
anomalously strong electric field. Nevertheless, it should be
stressed that these two different effects can simultaneously
co-exist in the presence of the fractal self-similarity on the
surface of nano-structured electrodes.

Therefore, the main goal of the paper was to develop an analytical
method of calculation of the giant local enhancement of the
optical electric field inside the fractal composite structure. It
should be admitted that nano-structured noble-metal-dielectric
composites are a standard object of nanoplasmonics. In our
consideration this is a nano-structured electrode - electrolyte
interface. In this
sense, we speak about nanoplasmonic electrochemistry. Owing to the
analytical expressions of enhanced electric field, one can analyze
the experimental results in the field of photo and
electrochemistry \cite{Cronin}. It is worth mentioning that all
methods of the calculation of the enhancement of the electric
field by fractal clusters used so for are different approximations
\cite{Stock1}.

We developed an analytical
approach, for description of the wave
propagation-localization in metal-dielectric nanostructures in the
\textit{quasi-static} limit. The method is based on fractional
calculus and permits to obtain an analytical expressions for the
electric field enhancement. This approach establishes a link
between fractional geometry of the nanostructure and fractional
integro-differentiation. An essential (geometrical) enhancement of
the electric field is obtained for the surface plasmon resonance
at a certain relation between permittivities of the host and
fractal metallic nanostructure, when ${\rm
Re}\vep_m(\omega)=-2\vep_d$ with a suitable detuning.

Important part of the analysis is developing convolution integrals
that makes it possible to treat the fractal structure. The initial
Maxwell equation (\ref{qsef1}) is local, since $l_0/\lambda\ll 1$
and space heterogeneity is accounted locally by virtue of the
characteristic function. An accurate treatment of the fractal
boundaries and recasting the Maxwell equation in the form of the
convolution integrals by accounting fractal properties of the
composite, eventually, leads to the coarse graining equations,
which already take into account the space heterogeneity and
nonlocal nature of the electric field and polarized dipole charges
inside the composite. Therefore, the heterogeneity, caused by the
fractal geometry, is reflected by the convolution of the averaged
fractal density and the electric field, according
Eqs.(\ref{theorRen}) and (\ref{qsef6}). It is necessary to admit
that this transform from ``local'' quasi-electrostatics to the
nonlocal, which takes into account space dispersion of
permittivity $\vep(r)$ is mathematically justified and rigorous
enough \cite{Ren}. The obtained convolution integrals are averaged
values, since the fractal density $r^{D-3}$ is the averaged
characteristics of the fractal structure, and, eventually, it
determines the space dispersion of the permittivity $\vep(r)$ of
the mixture.

Summarizing, we have to admit that observation of macroscopic
Maxwell's equations is related to averaging of microscopic
equations \cite{LLVIII}. This procedure for fractal composite
media is not well defined so far, since, according fractal's
definition, averaging over any finite volume depends on the size
of this volume itself \cite{falconer}. The main idea to overcome
this obstacle is to refuse the local properties of equations and
obtain nonlocal coarse graining Maxwell's equations, which are
already averaged.

\section*{Acknowledgments} We thank G. Zilberstein for helpful
discussions. This research was supported in part by the Israel
Science Foundation (ISF) and by the US-Israel Binational Science
Foundation (BSF).

\section{Appendix A: Fractional calculus briefly}

Extended reviews of fractional calculus can be found
\textit{e.g.}, in \cite{oldham,SKM,podlubny}. Fractional
integration of the order of $\alpha$ is defined by the operator
$${}_aI_x^{\alpha}f(x)=
\frac{1}{\Gamma(\alpha)}\int_a^xf(y)(x-y)^{\alpha-1}dy\, , $$
where $\alpha>0,~x>a$. Fractional derivation was developed as a
generalization of integer order derivatives and is defined as the
inverse operation to the fractional integral. Therefore, the
fractional derivative is defined as the inverse operator to
${}_aI_x^{\alpha}$, namely $
{}_aD_x^{\alpha}f(x)={}_aI_x^{-\alpha}f(x)$ and
${}_aI_x^{\alpha}={}_aD_x^{-\alpha}$. Its explicit form is
$${}_aD_x^{\alpha}f(x)=
\frac{1}{\Gamma(-\alpha)}\int_a^xf(y)(x-y)^{-1-\alpha}dy\, . $$
For arbitrary $\alpha>0$ this integral diverges, and, as a result
of this, a regularization procedure is introduced with two
alternative definitions of ${}_aD_x^{\alpha}$. For an integer $n$
defined as $n-1<\alpha<n$, one obtains the Riemann-Liouville
fractional derivative of the form
\begin{equation}\label{mt1a}   %
{}_a^{RL}D_x^{\alpha}f(x)=\frac{d^n}{dx^n}{}_aI_x^{n-\alpha}f(x)\,
,
\end{equation}
and fractional derivative in the Caputo form
\begin{equation}\label{mt1b}  %
{}_a^CD_x^{\alpha}f(x)= {}_aI_x^{n-\alpha}\frac{d^n}{dx^n}f(x)\, .
\end{equation}  %
There is no constraint on the lower limit $a$. For example, when
$a=0$, one has
${}_0^{RL}D_x^{\alpha}x^{\beta}=\frac{x^{\beta-\alpha}
\Gamma(\beta+1)}{\Gamma(\beta+1-\alpha)}\, , ~\alpha,\beta>0$.
This fractional derivation with the fixed low limit is also called
the left fractional derivative. Another important property is
$D^{\alpha}I^{\beta}=I^{\beta-\alpha}$, where other indexes are
omitted for brevity's sake. A convolution rule for the Laplace
transform for $0<\alpha<1$
\begin{equation}\label{mt3}
\clL[{}I_x^{\alpha}f(x)]=s^{-\alpha}\tilde{f}(s)
\end{equation}
is commonly used in fractional calculus as well.

\end{document}